\documentclass[preprint]{aastex}

\newcommand\etal{{\it et al.}}

\newcommand\kms{~{\rm km~s^{-1}}}

\newcommand\pc{~{\rm pc}}

\def\cm3{~{\rm cm^{-3}}}

\slugcomment{draft of \today}
\shorttitle{Collision of an Outflow with a Cloud}
\shortauthors{Baek et al.}

\begin{document}

\title{Numerical Simulations of a Protostellar Outflow
       Colliding with a Dense Molecular Cloud}

\author{Chang Hyun Baek\altaffilmark{1,2}, Jongsoo Kim\altaffilmark{3},
        and Minho Choi\altaffilmark{3}}

\altaffiltext{1}{Astrophysical Research Center for the Structure and Evolution of the Cosmos (ARCSEC),
Sejong University, Seoul, 143-747 Korea; chbaek@pusan.ac.kr}
\altaffiltext{2}{National Astronomical Observatory of Japan,
                 2-21-1, Osawa, Mitaka, Tokyo, 181-8588 Japan}
\altaffiltext{3}{International Center for Astrophysics,
                 Korea Astronomy and Space Science Institute,
                 Daejeon, 305-348 Korea}

\begin{abstract}
High-resolution SiO observations of the NGC 1333 IRAS 4A star-forming region
showed a highly collimated outflow with a substantial deflection.
The deflection was suggested to be caused by the interactions
of the outflow and a dense cloud core.
To investigate the deflection process of protostellar outflows,
we have carried out three-dimensional hydrodynamic simulations
of the collision of an outflow with a dense cloud.
Assuming a power-law type density distribution of the obstructing cloud,
the numerical experiments show
that the deflection angle is mainly determined by the impact parameter
and the density contrast between the outflow and the cloud.
The deflection angle is, however,
relatively insensitive to the velocity of the outflow.
Using a numerical model with physical conditions
that are particularly suitable for the IRAS 4A system,
we produce a column-density image
and a position-velocity diagram along the outflow,
and they are consistent with the observations.
Based on our numerical simulations,
if we assume that the initial density and the velocity of the outflow
are $\sim 10 \cm3$ and $\sim 100 \kms$,
the densities of the dense core and ambient medium in the IRAS 4A system
are most likely to be $\sim 10^5 \cm3$ and $\sim 10^2 \cm3$, respectively.
We therefore demonstrate through numerical simulations
that the directional variability of the IRAS 4A outflow
can be explained reasonably well using the collision model.
\end{abstract}


\section{Introduction}

Radio interferometric observations in molecular lines
show directional variability of jets in some star forming regions.
The understanding of the variability of protostellar jets
provides us with clues to a variety of physical processes
involved in the interaction of jets and surrounding clouds.
To explain the variability of protostellar jet,
several models were proposed \citep[]{em97,gcr99,choi2001a}.
These models can be classified into two categories.
The first category includes external perturbation models,
such as density gradients in the environment,
flow instabilities (e.g., Kelvin-Helmholtz instability),
side winds, and magnetic fields.
The second category includes intrinsic variability models,
such as the motion of the jet source perpendicular to the jet direction
and the precession of the jet.
Moreover, to understand the bending of the jet,
especially in the context of young stellar objects (YSOs),
numerical simulations have been carried out by several authors
\citep[]{dp99,hurka99,rc95,raga2002}.
Three-dimensional smoothed particle hydrodynamic simulations
of the penetration of a jet into a dense stratified cloud
were carried out by \citet{dp99},
and it was found that a radius ratio and a density ratio
between the jet and the cloud
are important parameters in determining the outcome.
\citet{hurka99} studied non-radiative jets
bent by magnetic fields with a strong gradient.
They showed that a low-velocity jet can be deflected
by magnetic fields with a modest strength,
but a high-velocity jet needs very strong fields to be deflected.
\citet{rc95} presented analytical and two-dimensional numerical
studies of the collision of an Herbig-Halo jet
with a dense molecular cloud core.
\citet{raga2002} studied the interaction between a radiative jet
and a dense cloud through three-dimensional hydrodynamic simulations
and generated H$\alpha$ and H$_2$ $1-0$ S(1) emission maps.

The NGC 1333 molecular cloud contains numerous YSOs
and outflows and is a well-studied star formation region.
A highly collimated outflow with a substantial deflection angle
was observed from high resolution SiO observations
of NGC 1333 IRAS 4A \citep{choi2005a}.
To investigate the overall structure of the IRAS 4A outflow system,
\citet[]{choi2001b,choi2005a,choi2005b} and \citet{choi2006}
observed the outflow in the C$^{18}$O, $^{13}$CO $J=1 \rightarrow 0$,
HCO$^+$, HCN $J=1 \rightarrow 0$,
SiO $v=0$ $J=1 \rightarrow 0$, and H$_{2}$ $1-0$ S(1) lines.
\citet{choi2005a} suggested
that the sharp bend was caused by a collision between
the northeastern outflow and a dense core in the ambient molecular cloud.
And then, \citet{choi2005b} confirmed that the obstructing cloud
with a dense core is located just north of the bending point
in images of the HCO$^+$ and HCN $J=1 \rightarrow 0$ lines.
A sharp bending of the northeastern outflow in the NGC 1333 IRAS 4A system
is unusual structure and
provides a good example to study
the bending mechanism of outflows in protostellar environment.

In this study we have carried out three-dimensional hydrodynamic simulations
on the interactions of an outflow with a dense molecular cloud
surrounded by an ambient medium,
in order to understand the deflection mechanism of the outflow
and to construct a detailed model of deflected outflows.
We performed many simulations with different sets of physical parameters
that can change the outflow direction,
then picked up one model that gives a similar deflection angle
with the observed one in the IRAS 4A system.
Finally, we make a column-density image of shocked gas
and a position-velocity diagram along the outflow.
They are reasonably consistent with the observed data
of the IRAS 4A northeastern outflow.
In \S 2, we describe a numerical setup
for the interaction of an outflow and a cloud.
Simulation results are presented in \S 3,
followed by the summary and discussion in \S 4.

\section{Numerical Simulations}

We execute three-dimensional hydrodynamic simulations
of outflow/cloud interactions
with a grid-based hydrodynamic code based on the total variation diminishing (TVD) scheme \citep{ryu93}.
We integrate numerically the hydrodynamic equations,
\begin{equation}
{\partial \rho \over \partial t} + \nabla \cdot (\rho \mathbf v) = 0,
\end{equation}
\begin{equation}
{\partial \over \partial t} (\rho \mathbf v)
+ \nabla \cdot (\rho \mathbf v \mathbf v + p \mathbf I) = 0,
\end{equation}
\begin{equation}
{\partial E \over \partial t} + \nabla \cdot [ (E + p) \mathbf v] = 0,
\end{equation}
where $E = (1/2) \rho \mathbf v^2 + p/(\gamma-1)$
and other notations have their usual meanings.
For the adiabatic index, $\gamma=5/3$ is assumed.
We did not take into account radiative cooling and heating processes, and chemical reactions, except for a test run explained below.
The reason why we ignored them is that the deflection angle,
which is of our main interest, is not sensitive to them.
Two numerical simulations carried out by \citet{dp99} showed
that the deflection angle ($\theta \simeq 40 \arcdeg$)
seen in the adiabatic simulation
is slightly larger than the angle ($\theta \simeq 35 \arcdeg$)
in the simulation with radiative cooling.
To check their result,
we also performed two simulations with and without radiative cooling
\citep[]{suth93,san2002} and heating \citep{bkkr05} processes.
The structure of the outflow before and after collision in our two simulations are similar,
except that a denser and shallower shell
is formed at the head of the deflected outflow in the simulation with the radiative processes.
The deflection angle shown in the adiabatic simulation is about $\sim 3 \arcdeg$ larger than
that shown in the simulation with the radiative processes.
Both the results of \citet{dp99} and ours suggest
that the radiative cooling
does not play a major role in the outflow deflection.

We setup a computational box whose minimum and maximum Cartesian coordinates are defined by
$-0.05\pc <x< 0$, $-0.025\pc <y< 0.025\pc$, and $-0.04\pc <z< 0.06\pc$.
The box is covered with $256 \times 256 \times 512$ cells, so the numerical resolution in each direction becomes a constant, $40$ AU.
A cloud with a radius $R_{c}=0.02 \pc$ is located at the origin of the coordinate system. An outflow with a sectional radius $R_{o}=300$ AU is injected from the bottom $x-y$ plane into the box with a direction parallel to the $z$-axis.
Open boundary conditions are used on all boundaries,
except for the bottom of $x-y$ plane, on which a reflection condition is imposed.
To explore the outflow/cloud interactions with different physical parameters,
we vary the density distribution of a molecular cloud
(an initially spherical molecular cloud with uniform
density or power-law type density distribution),
an impact parameter (from $3/10 R_c$ to $10/10R_c$),
and the density ratios among the outflow, cloud, and ambient medium.
The initial ratios of the number density of outflow gas $n_{o}$
to the number density of the ambient gas $n_{a}$
and the number density of outflow gas to the number density of cloud gas $n_{c}$
are defined as, respectively
\begin{equation}
\eta={n_{o} \over n_{a}},
\chi={n_{o} \over n_{c}}.
\end{equation}
Twelve simulations were done with different combinations of the model parameters, which are summarized in Table~1.

\section{Results}

\subsection{Uniform Density Cloud Models}

To compare our simulation results with the results of \citet{raga2002},
we performed numerical experiments on the interaction of an outflow with a uniform density cloud
for different impact parameters, $4/10 R_{c}$, $6/10 R_{c}$, $8/10 R_{c}$,
and $10/10 R_{c}$ (See Table 1).
The outflow with number density $n_{o}=10 \cm3$,
temperature $T_{o}=10^4$ K, and velocity $100 \kms$
is injected toward an initially spherical cloud with uniform
density $n_{c}=10^3 \cm3$.
The spherical cloud is surrounded by a homogeneous ambient
medium of density $n_{a}=1 \cm3$ and temperature $T_{a}=10^4$ K.
The cloud is in pressure equilibrium with the ambient medium.
Figure 1 shows density and velocity fields
on the $y=0$ plane for models UC01, UC02, UC03, and UC04.
At $t=200$ yr, the outflow has reached the cloud boundary and bores a hole through the cloud or gets deflected.
In the case of the models with small impact parameters (UC01 and UC02)
the incident outflow impinges on the surface of the cloud, makes a hole,
and finally penetrates through the cloud.
In the other case of the models with large impact parameters (UC03 and UC04),
the incident outflow is deflected by the cloud,
and then the deflected flow propagates with a wider opening angle.
Depending on the impact parameter,
the evolution of the interaction of an outflow with a uniform cloud
can be classified into two cases as explained above.
For the case of deflected outflow,
the impact parameter is one of the important parameters
that determines the deflection angle,
which is the angle between the incident outflow direction
and the deflected flow direction.
Our results of the two uniform cloud models UC02 and UC03
are basically similar to those of model A of \citet{raga2002}.

\subsection{Power-Law Type Cloud Models}

As we have pointed out in the introduction section,
the main motivation of this paper
is to make a numerical model of the deflected outflow
seen in the NGC 1333 IRAS 4A system.
So, whenever possible, we pick up physical quantities such as number density, and temperature  from observations of the system.  If not, we resort to the known quantities for the typical protostellar outflows and their environment.
An outflow with number density $n_{o}=10 \cm3$
and temperature $T_{o}=10^4$ K
propagates through the uniform ambient medium
with number density $n_{a}=100 \cm3$ and temperature $T_{a}=10$ K.
The outflow interacts with an obstructing molecular cloud
with a power-law type density profile,
\begin{equation}
n_{cl} =  {{n_{c}} \over {[1+A(r-r_{c})]^2}}
~~(R_{c} \geq r \geq r_{c}),
\end{equation}
where the distance from the cloud center $r = \sqrt{x^2+y^2 +z^2}$,
the cloud radius $R_{c}=0.02 \pc$, the core radius $r_{c}=400$ AU,
and the scaling factor $A \sim 17$ AU$^{-1}$.
The number density and temperature in the center of the dense core
are $n_{c}=10^5 \cm3$ and $T_{c}=10$ K, respectively.  With the given number densities and temperatures for the outflow, cloud, and ambient medium, there is no pressure balance among those gas components.
However, considering the fact that there is one order of magnitude difference
between the largest sound speed with temperature 10$^{4}$ K
and an outflow speed 100 km s$^{-1}$,
the motion induced by the pressure imbalance is negligible.

Figure 2 depicts the density and velocity fields on the $y=0$ plane
for models PC01, PC02, PC03, and PC04.
For uniform density cloud models with smaller impact parameters (UC01 and UC02),
the outflow makes a hole into and through the cloud (see Figure 1).
For the power-law type density models, on the contrary,
the outflow with a small impact parameter
does not make a hole through the cloud but is deflected significantly.
Furthermore, the deflection angle of the outflow increases as the impact parameter decreases.  This result demonstrates that not only the impact parameter but also the density gradient of the cloud plays an important role in determining the deflection angle.
We also measure deflection angles as a function of time
from the simulation results of models PC01, PC02, PC03, and PC04, which is shown in
Figure 3.
The increase of deflection angles in the cases of small impact parameters (models PC01 and PC02)
is quite significant, which is due to large density contrast between the outflow and the inner part of the obstructing cloud.  
It seems to us that, when the impact parameter is smaller than a critical value, 
the deflection angle increases with time and then keeps its angle after passing over the cloud.
But outflow in models PC03 and PC04
is not efficiently disturbed by the obstructing cloud.
For models PC03 and PC04 the deflection angles are small,
because the outflow interacts with the outer low density part of the cloud.
After being deflected by the cloud,
the outflow again keeps its small deflection angles
for these models with large impact parameters.

Figure~4 shows the time evolution of our fiducial model PC02, whose model parameters are
$n_{o}=10 \cm3$, $\eta=0.1$, and $\chi=10^{-4}$ (see Table~1).
At $t=700$ yr, a bow shock generated by the outflow reaches the cloud boundary,
and the velocity at the bow shock front is slightly changed
due to the density gradient of the cloud boundary.
The propagation direction of the outflow changes only slightly
when the bow shock just passes through the cloud boundary.
The larger deflection occurs
at the nearest point from the center of the cloud at $t \approx 1800$ yr,
and then the deflection angle tends to increase with time
as the outflow passes over the cloud.
At later times ($t\gtrsim 4200$ yr),
the deflected outflow moves into the ambient medium
without any significant change of deflection angle.
It is known that the outflow velocities
before and after the collision with the cloud
has a relation \citep{rc95},
\begin{equation}
v_d \approx v_i \cos \theta,
\end{equation}
where $v_d$ is the velocity of the deflected outflow, $v_i$ is the velocity of the incident outflow,
and $\theta$ is the deflection angle.
At $t=4200$ yr,
a mean velocity of the deflected outflow $v_d \sim 80\kms$
measured from our simulation
is approximately equal to the value $v_d = 86.6\kms$
(for $v_i = 100 \kms$ and $\theta \simeq 30 \arcdeg$)
predicted by from the equation above.
The deflection angle after $t=3800$ yr for model PC02 is
very similar to the observed value of the northeastern outflow of IRAS 4A.

\subsection{Models with Different Density Contrast and Outflow Velocity}

The effects of the outflow density and outflow velocity on the deflection angle
can be examined by comparing the results from simulations with
different outflow densities (models PC11 and PC12) and velocities (models PC21 and PC22).
Models PC11 and PC12 have
hundred ($\eta=10$, $\chi=10^{-2}$) and ten ($\eta=1$, $\chi=10^{-3}$) times
larger outflow densities, respectively,
than that of PC02 ($\eta=0.1$, $\chi=10^{-4}$).
And models PC21 and PC22 have two ($200 \kms$) and three ($300 \kms$) times
larger outflow velocities, respectively,
than that of PC02 ($100 \kms$).
Figure 5 shows the density and velocity fields for models PC11, PC12, PC21, and PC22.
As shown in Figure 3, the light outflow,
whose number density is smaller
than the number density of ambient medium ($\eta \leq 0.1$),
experiences quite a large deflection,
and its opening angle after collision is small.
However the heavy outflow,
whose number density is larger than or equal
to the number density of ambient medium ($\eta \geq 1$),
experiences a slight deflection,
and then it propagates with a wider opening angle
(see Fig.~5 for models PC11 and PC12).
The kinematic time scale, which is defined by the time required for the outflow to cross over the longer dimension of the computational box, is dependent upon the $\eta$ parameter.
In fact, the lower-left two panels of Figure~5 show
that a bow shock of the model PC11
arrives the upper boundary at $t=2000$ yr,
whereas the bow shock of the model PC21 does at $t=3500$ yr.
This is because, as the density of the outflow increases with respect to the density of the ambient medium, the outflow is less decelerated by the medium.  The lower-right two panels show that the overall density and velocity structures of them are very similar to each other except for the different time epochs.  The evolutionary time of the faster outflow model PC22 should naturally be shorter than that of model PC21.  This result show that the outflow velocity does not influence the deflection angle.
Considering the three observational facts of the IRAS 4A outflow,
(i) the deflection angle of $\sim 30\arcdeg$, (ii) the well-collimated flow,
and (iii) the kinematic time scale of the outflow,
PC02 is the fiducial model that describes the IRAS 4A system most closely.

\subsection{Comparison between Observational and Simulated Results}

The deflection angle of the outflow in model PC02 may be compared directly
with the observed angle of the IRAS 4A SiO outflow (Figure 6).
To make the simulated outflow image shown in the left box of Figure~6,
we first pick up shocked gas whose temperature is higher than 3 times the initial outflow temperature ($T > 3T_{o}$), then integrate the shocked gas along the $y$-axis.
We can find, in the simulated outflow image,
the well collimated flow and the enhanced column density
at the bending point and the terminal part of the outflow,
which are very similar to the observed SiO image.
Moreover, the time epoch of the simulated outflow image
is similar to the kinematic timescale of IRAS 4A outflow determined by
the SiO main outflow observation \citep{choi2005a}.

The origin of the SiO molecules is not well-known.
They can be either injected into the flow
at the base of the outflow near the protostar
or gradually mixed in from the ambient cloud.
To trace the outflow and cloud separately,
we have implemented a method of the Largrangian tracer \citep{tj96} into our Eulerian hydrodynamic code.
From this Largrangian tracer,
we find that the shocked gas mainly consists of material from the outflow,
although a small amount of material from the cloud and ambient medium
is entrained in the deflected shock gas.
From the information on the amount of shocked gas, we infer that most of the observed SiO gas might be from the outflow.
However, our statement needs to be confirmed
by simulations that explicitly includes the chemistry of SiO
and related species.

The complicated kinematics of the deflected flow
was revealed using position-velocity diagram.
Figure 7 depicts the column density image
and the position-velocity diagram
along the solid line in the column density image calculated from model PC02.
When we make the position-velocity diagram,
we consider that the $y=0$ plane coincides with the plane of the sky
and take into account the inclination angle $9 \degr$ of the outflow
with respective to the plane of the sky,
which was chosen from two assumptions ; 1) the typical velocity of the outflow is $100 \kms$, 
2) the velocity interval in the position-velocity diagram is $\sim 20 \kms$.
The position-velocity diagram shows that the line-of-slight velocity of the deflected part is slower and more turbulent than that of the undeflected part.
This implies that the outflow is slowed down and heated
by the collision with a dense molecular core and the interaction with the ambient medium.
The broad-velocity structure near the head of the deflected outflow
is similar to that of the position-velocity diagram of the SiO observation.
It was suggested \citep{mk04} and investigated \citep{nl07}
that outflows might be one of the mechanisms
that drive turbulence in molecular clouds.
In our simulations, even though they are single collision experiments of the outflow and the cloud, 
we demonstrate that a turbulent flow with a velocity dispersion of the order of 10 km sec$^{-1}$ can be easily generated.

\section{Summary and Discussion}

We have performed three-dimensional hydrodynamics simulations
to study the interaction of an outflow with a dense molecular cloud.
Since the most evident information from observations is the deflection angle,
we focus on the deflection angle in the simulations by varying four parameters:
the impact parameter, the outflow density contrast with the ambient medium ($\eta$),
the contrast between outflow and cloud density ($\chi$),
and the velocity of outflow.
The main results can be summarized as follows.
\begin{itemize}
\item When the impact parameter is smaller than 0.6 times the cloud radius, the outflow bores a hole into the uniform cloud.  On the contrary, the outflow with the impact parameter as small as 0.3 is deflected by the cloud with the power-law type density distribution.
\item When the obstructing cloud has a power-law density distribution,
      the deflection angle of the outflow
      is determined by the impact parameter
      and the density contrast between the outflow and the cloud ($\chi$).
\item The deflection angle is relatively insensitive
      to the velocity of outflow.
\item The initially light outflow ($\eta<1$) can be more collimated than the heavy one ($\eta>1$).
\end{itemize}

\citet{choi2005a} suggested
that the NGC 1333 IRAS 4A outflow was deflected
as a result of a collision with a dense cloud core
and provided several lines of evidence.
They are
(1) the asymmetric morphology of the bipolar outflow,
(2) the good collimation and complicated kinematics of the deflected flow,
(3) the low-velocity emission from the molecular gas near the bend,
and (4) the enhancement of SiO emission in the deflected flow.
In this study, we confirmed these evidences through numerical experiments.
Especially, the deflection angle ($\sim 30\arcdeg$) of collimated outflow
and the enhancement and complicated kinematics of shocked gas
in the deflected region produced from our numerical simulations
are very similar to those seen in the SiO observations of the IRAS 4A region.
If we take model PC02
as the most appropriate numerical model for the IRAS 4A outflow,
we can infer that this outflow was significantly deflected
by the dense molecular core $\sim 2000$ years ago
and the ratio of the outflow density to the density at the cloud center
is very low $\sim 10^{-4}$.
Moreover,
if we assume that the initial density and the velocity of the outflow
are $\sim 10 \cm3$ and $\sim 100 \kms$,
the densities of the dense core and ambient medium in the IRAS 4A system
are most likely to be $\sim 10^5 \cm3$ and $\sim 10^2 \cm3$, respectively.

Although our simulations do not consider
the radiative cooling and chemical reactions,
our results provide some insights
to the interactions of an outflow with a dense cloud.
To make a more sophisticated numerical model
compatible with the SiO observations,
we need to include chemistry, especially related to the SiO molecule.
In the near future,
we will execute numerical simulations
which include the radiative cooling
and chemical reactions related to the SiO molecule
and then try to directly compare the simulated SiO emission image
with the observed one.

\acknowledgments

We thank H. Kang for supporting her advection code.
C.H.B would like to thank K. Tomisaka and T. Kudoh for valuable discussions.
Simulations presented in this paper were done using VPP5000 at the National Astronomical Observatory of Japan (NAOJ) and a high performance cluster that was built
with the funding from the Korea Astronomy and Space Science
Institute (KASI) and the Astrophysical Research Center for
the Structure and Evolution of the Cosmos (ARCSEC) of the
Korea Science and Engineering Foundation (KOSEF).
The work of  C.H.B. was supported by the Korea Research Foundation
Grant funded by the Korean Government (KRF-2006-352-C00030).
The work of J.K. was supported by KOSEF through ARCSEC
and the grant of the basic research program R01-2007-000-20196-0.

\clearpage
\begin{deluxetable}{cccccccccccccccc}
\tablewidth{0pc}
\tablecaption{Model Parameters for Simulations}
\tablehead{\colhead{Model}
& \colhead{Impact Parameter}
& \colhead{Cloud Density Distribution}
& \colhead{$\eta = {n_{o} \over n_{a}}$}
& \colhead{$\chi = {n_{o} \over n_{c}}$\tablenotemark{a}}
& \colhead{$v$\tablenotemark{b}} }
\startdata
UC01 &  4/10 $ R_{c}$ & uniform        & 10  & $10^{-2}$ & 100\\
UC02 &  6/10 $ R_{c}$ & uniform        & 10  & $10^{-2}$ & 100\\
UC03 &  8/10 $ R_{c}$ & uniform        & 10  & $10^{-2}$ & 100\\
UC04 & 10/10 $ R_{c}$ & uniform        & 10  & $10^{-2}$ & 100\\
PC01 &  3/10 $ R_{c}$ & power-law type & 0.1 & $10^{-4}$ & 100\\
PC02 &  4/10 $ R_{c}$ & power-law type & 0.1 & $10^{-4}$ & 100\\
PC03 &  5/10 $ R_{c}$ & power-law type & 0.1 & $10^{-4}$ & 100\\
PC04 &  6/10 $ R_{c}$ & power-law type & 0.1 & $10^{-4}$ & 100\\
PC11 &  4/10 $ R_{c}$ & power-law type & 10  & $10^{-2}$ & 100\\
PC12 &  4/10 $ R_{c}$ & power-law type & 1   & $10^{-3}$ & 100\\
PC21 &  4/10 $ R_{c}$ & power-law type & 0.1 & $10^{-4}$ & 200\\
PC22 &  4/10 $ R_{c}$ & power-law type & 0.1 & $10^{-4}$ & 300\\
\enddata
\tablenotetext{a}{For the power-law type density models, $n_{c}$
represents the central number density of a cloud.}
\tablenotetext{b}{The outflow velocity in units of $\kms$.}
\end{deluxetable}

\clearpage
\begin{figure}
\plotone{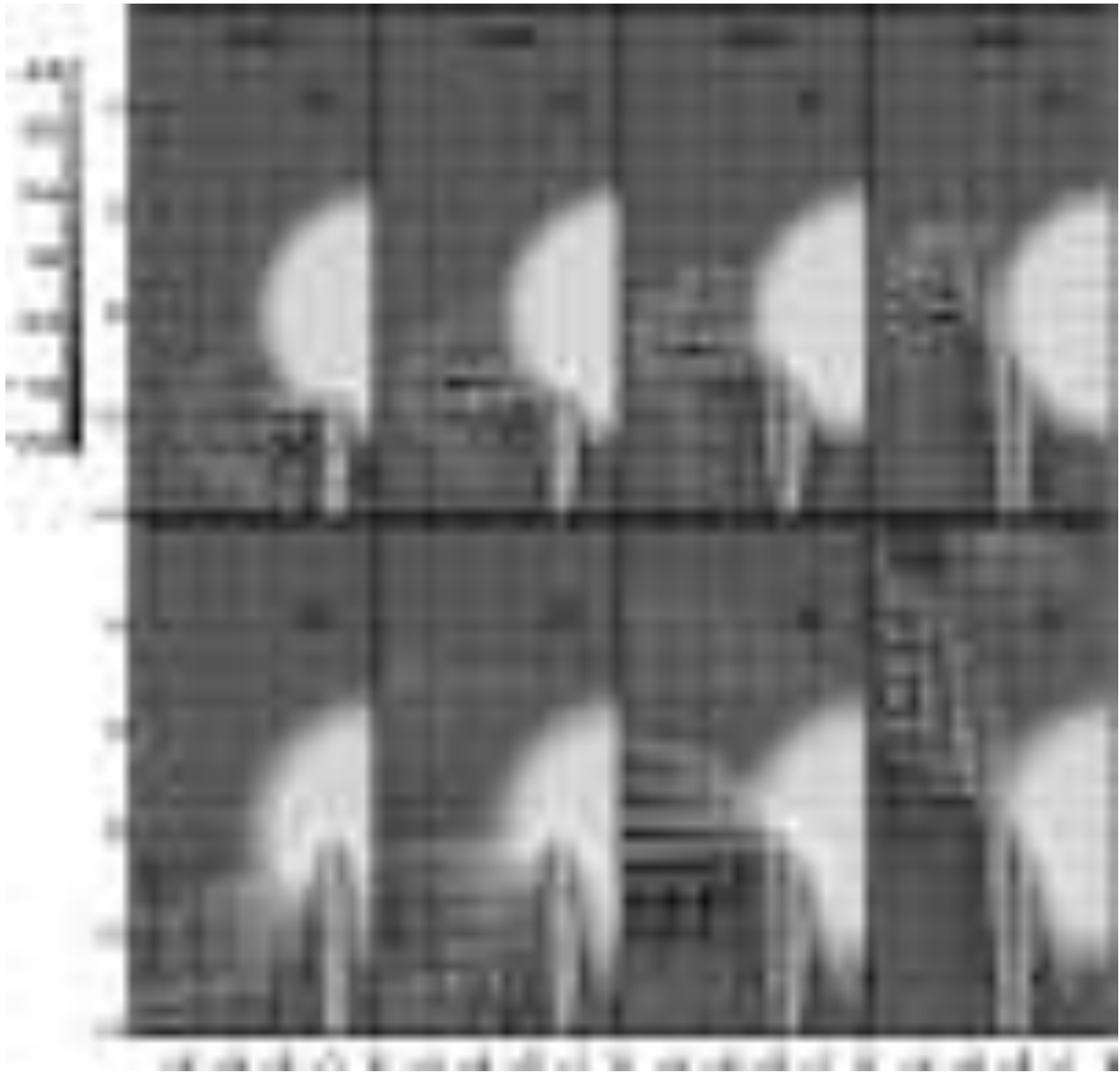}
\figcaption
{Density and velocity fields of uniform density cloud models
UC01, UC02, UC03, and UC04.
Gray images for density fields and arrows for velocity fields on the $y=0$ plane
(which includes the outflow axis and the center of the spherical cloud)
are shown in the upper panels at a time epoch $t=1000$ yr and
the lower ones at $t=3000$ yr.
The gray scale bar is in logarithmic units.
The longest arrow corresponds to $100\kms$.}
\end{figure}

\begin{figure}
\plotone{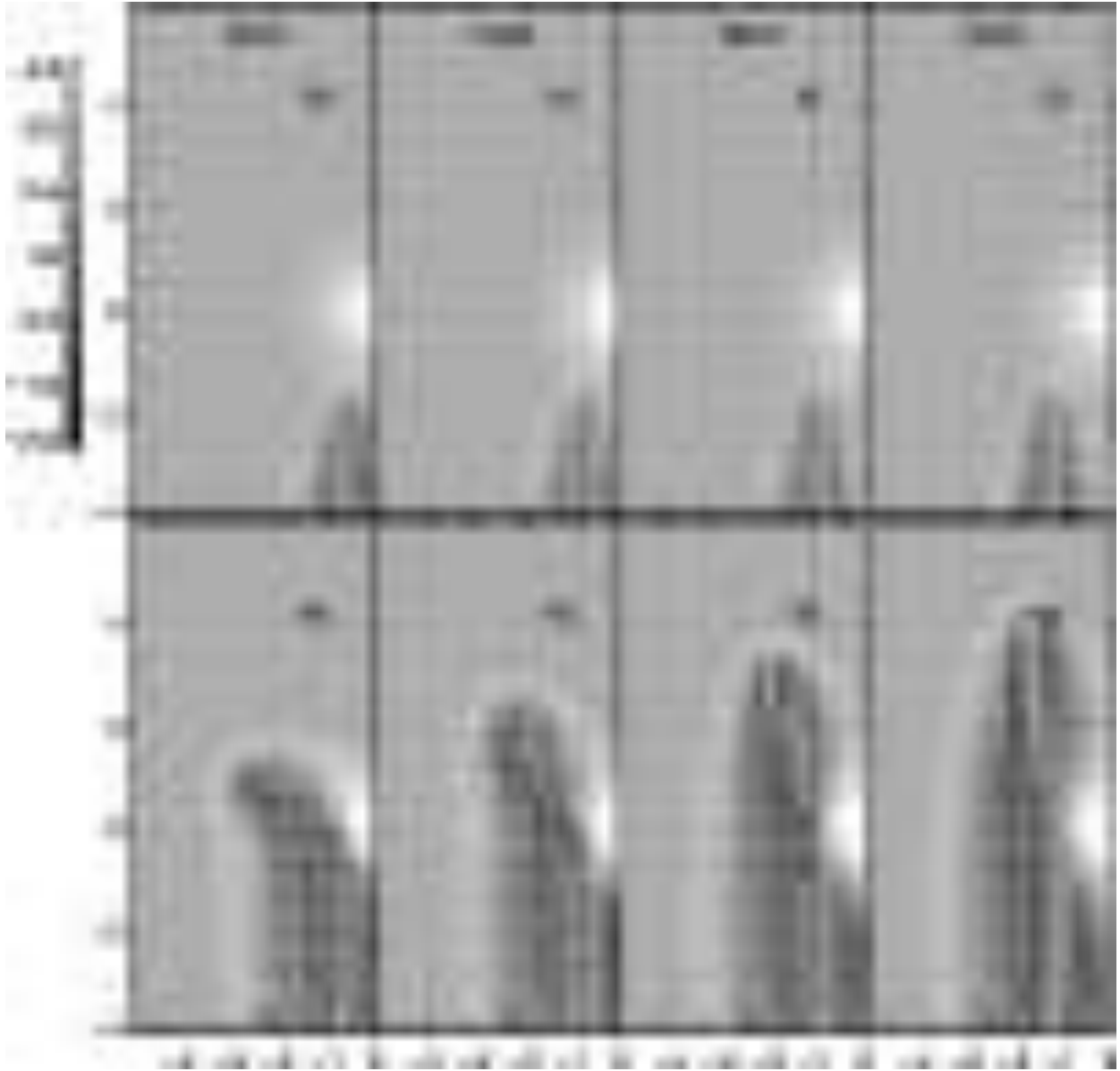}
\figcaption
{Density and velocity fields of power-law type density cloud models
PC01, PC02, PC03, and PC04.
The model name is given at top of each upper panel.
The upper and lower panels
show gray images for density fields and arrows for velocity fields
at time epochs $1000$ yr and $4500$ yr, respectively.
The gray scale bar is in logarithmic units.
The longest arrow corresponds to $100 \kms$.}
\end{figure}

\begin{figure}
\plotone{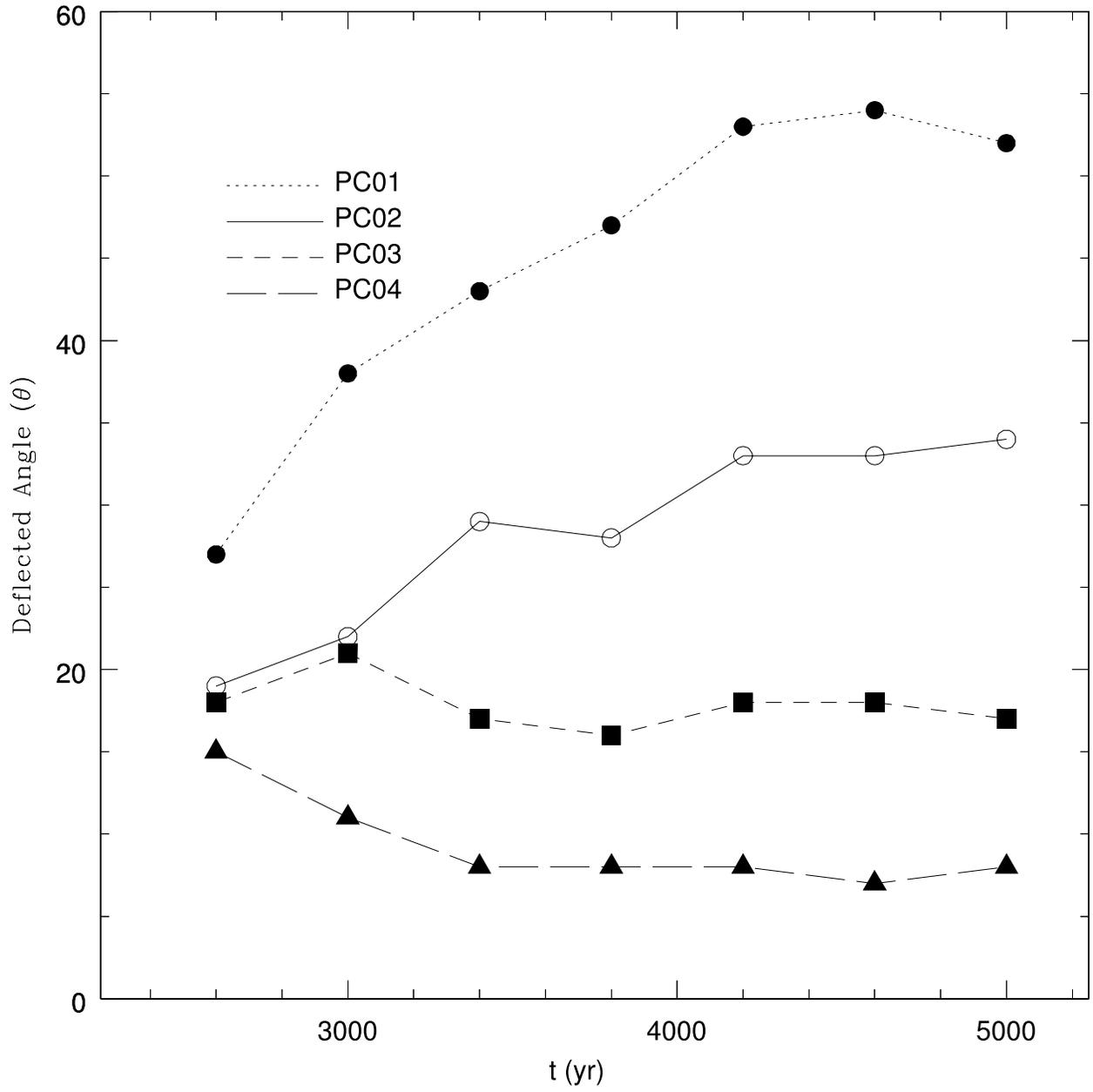}
\figcaption
{Evolution of deflection angle for models
PC01 (dotted line; filled circle), PC02 (solid line; open circle),
PC03 (dashed line; filled box), and PC04 (long-dashed line; filled triangle).
}
\end{figure}

\begin{figure}
\plotone{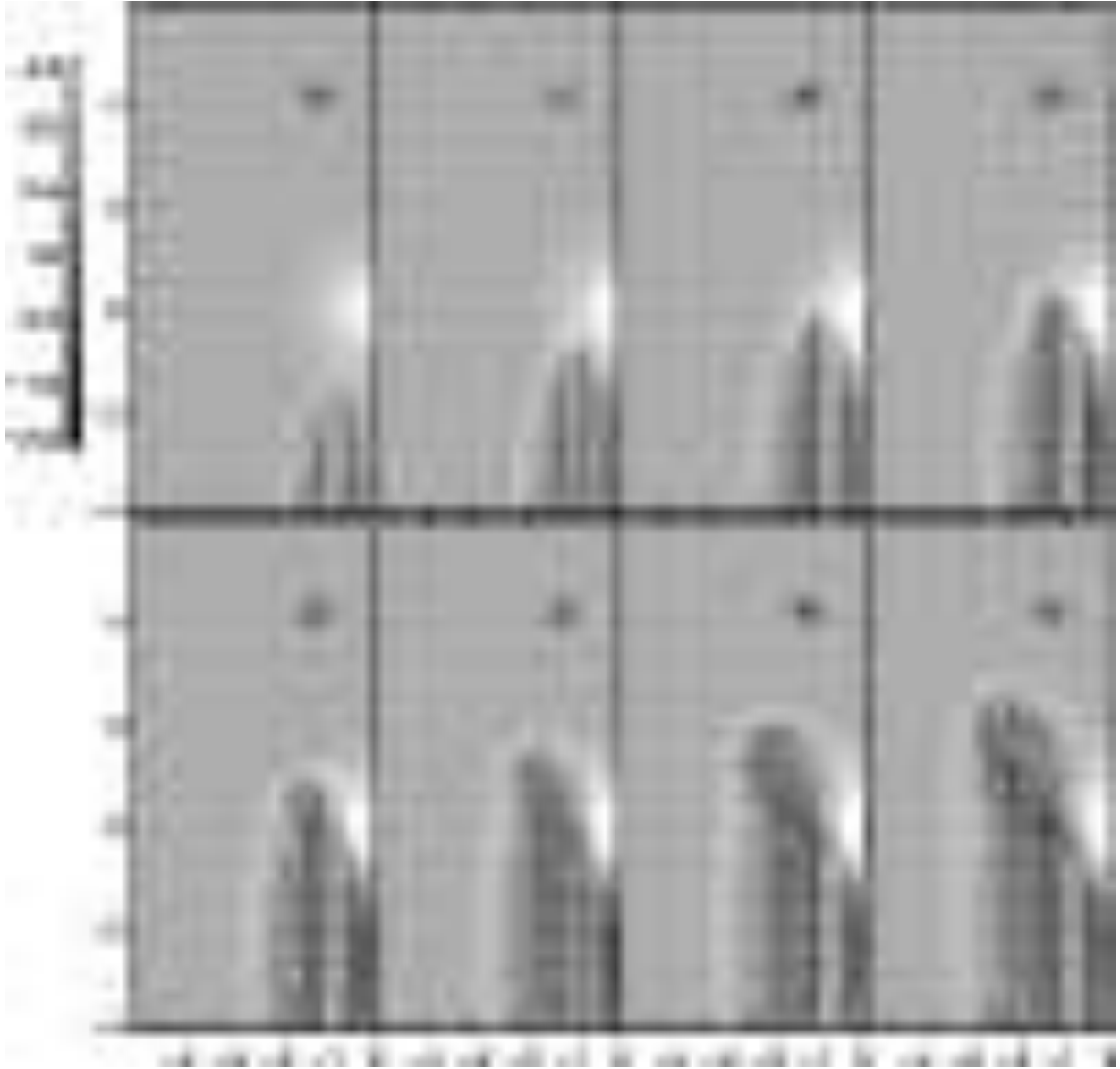}
\figcaption
{The time evolution of density and velocity fields for model PC02. Gray images for density fields and arrows for velocity fields on the $y=0$ plane are shown.  A two digit number in each panel shows a time epoch in units of $100$ yr.
}
\end{figure}

\begin{figure}
\plotone{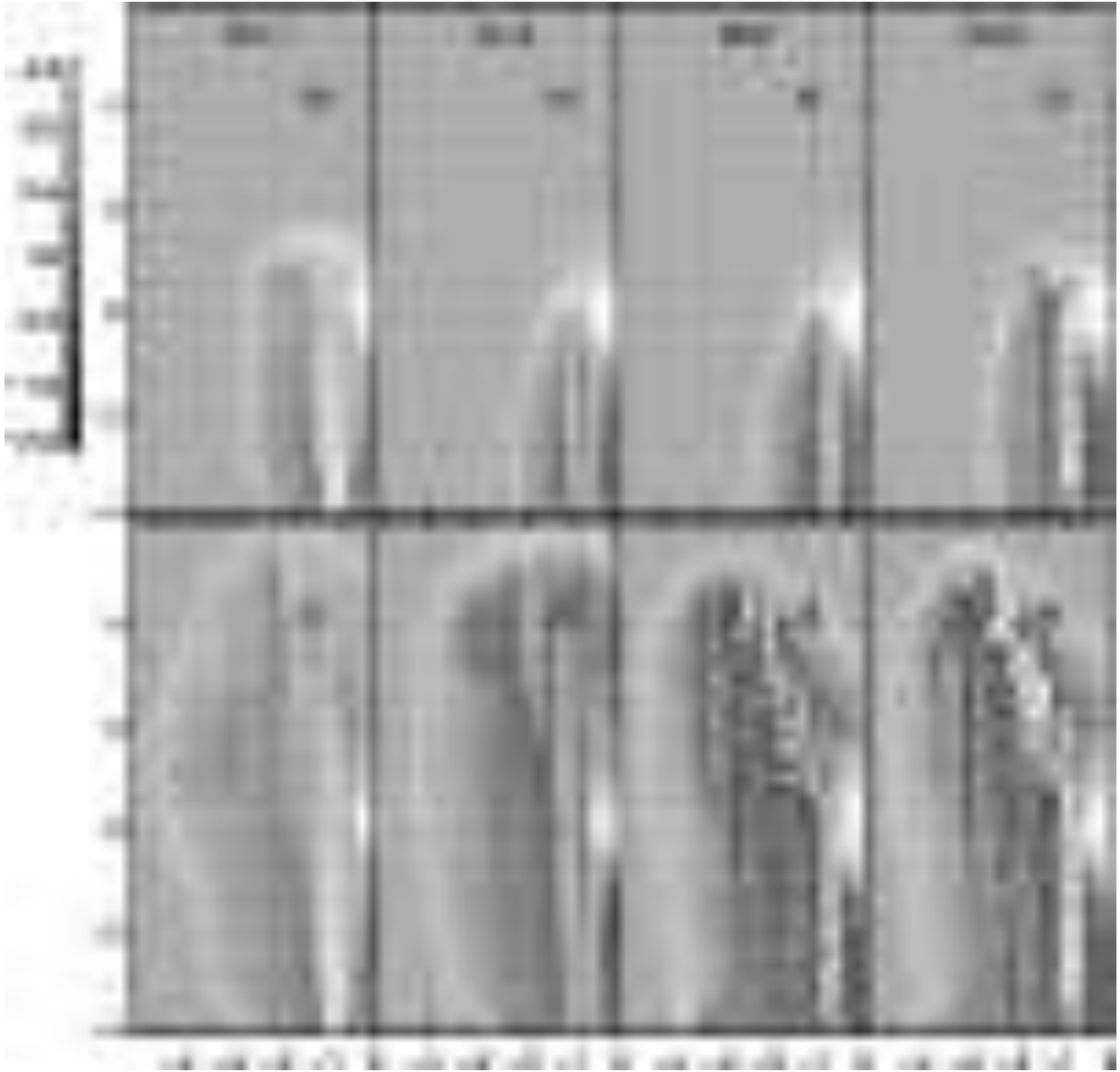}
\figcaption
{Density and velocity fields for models PC11, PC12, PC21, and PC22.
Gray images for density fields and arrows for velocity fields on the $y=0$ plane are shown. A gray scale bar in logarithmic units is included in the far left side.  The longest arrow corresponds to a speed around $300\kms$. A two digit number in each panel shows a time epoch in units of $100$ yr.
}
\end{figure}

\begin{figure}
\plotone{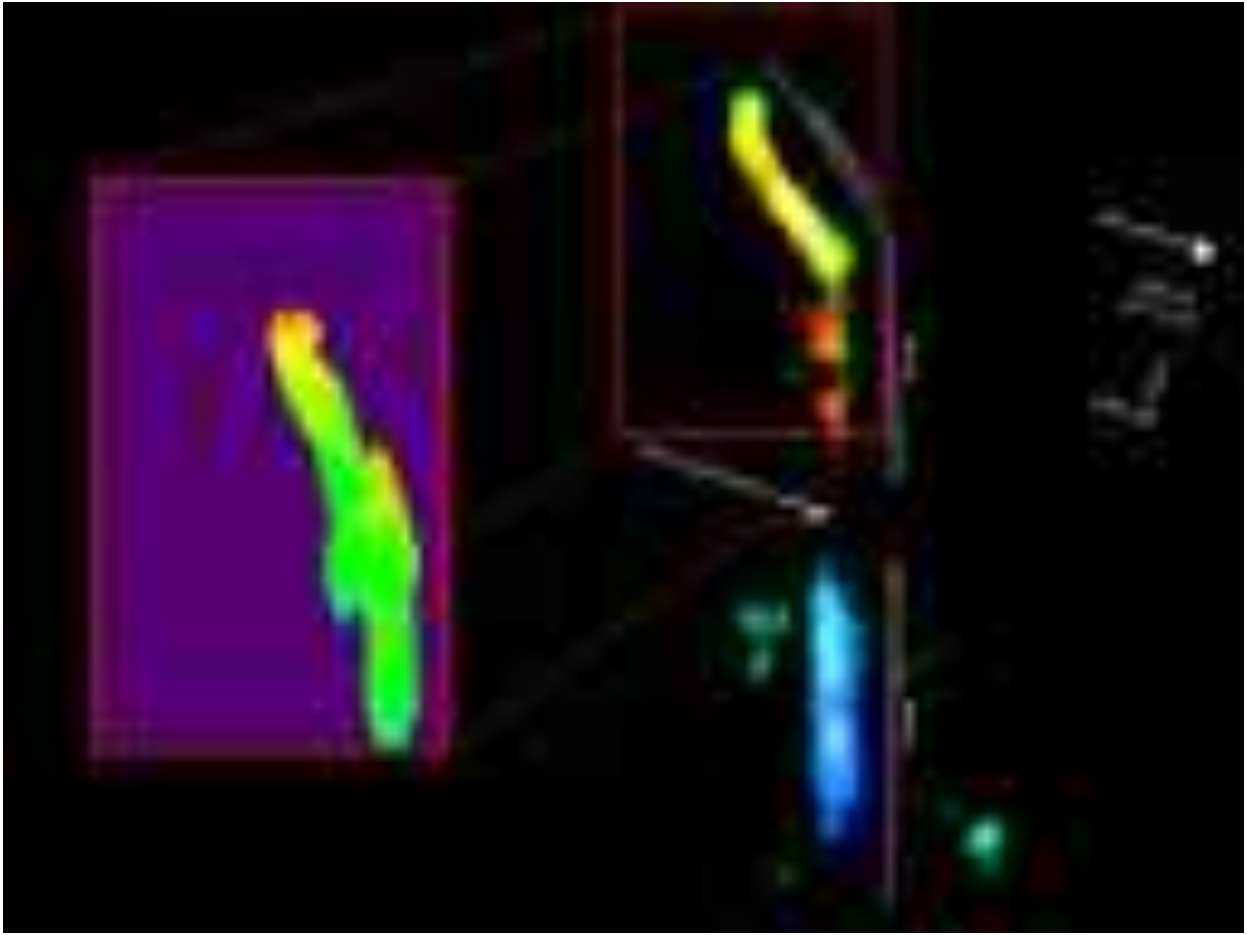}
\figcaption
{Comparison of observed and simulated outflows \citep[from][]{bkc07}.
The right box contains a SiO image
toward the northeastern region of NGC 1333 IRAS 4A.
The left box shows
an image of the column density of shocked gas
at a time epoch of $4200$ yr of the numerical model PC02.
The deflection angles seen in both boxes are very similar.
}
\end{figure}

\begin{figure}
\plotone{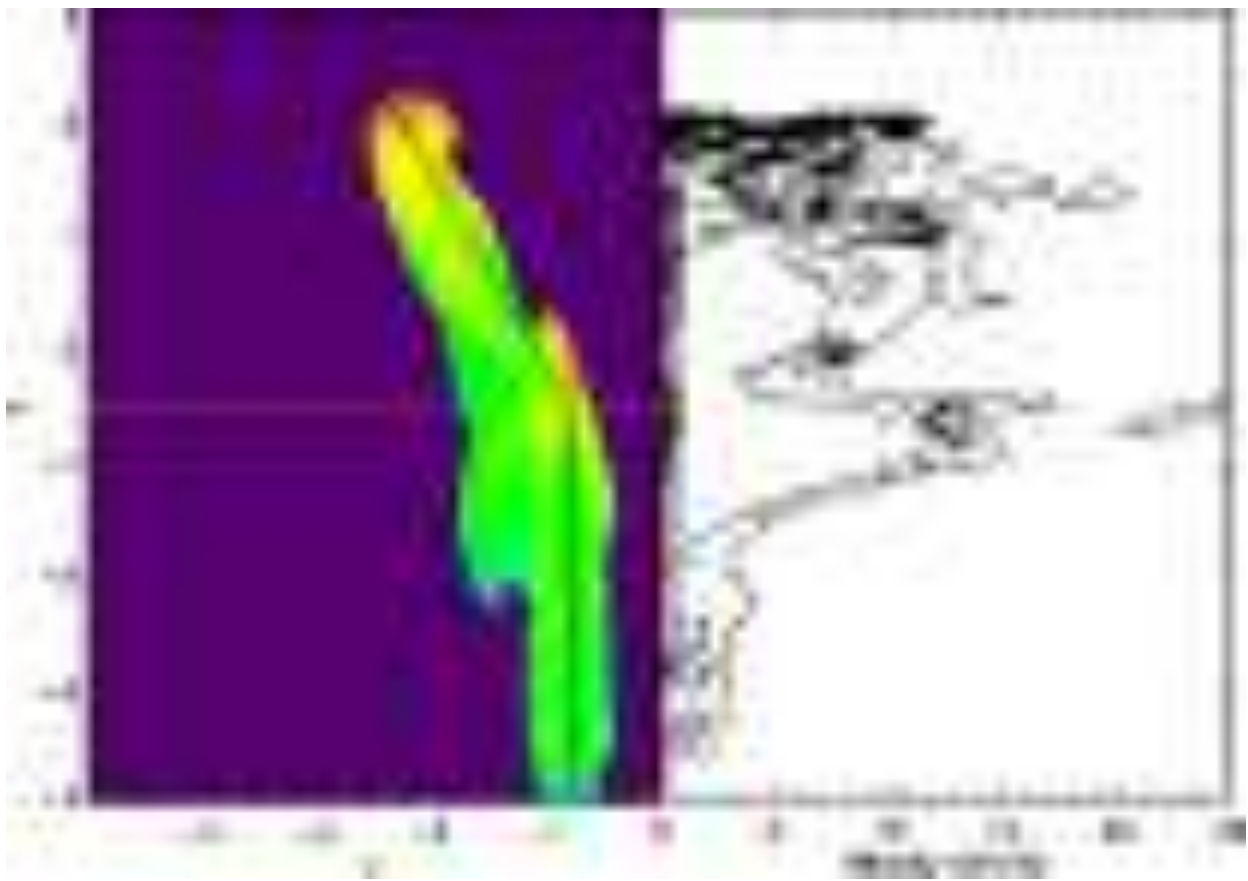}
\figcaption
{The column-density image and position-velocity diagram
of model PC02 in Table~1.
The position-velocity diagram in the right panel is plotted
along the solid line in the left panel.
The intersection of the solid and dashed lines in the left panel corresponds to the bending point of the outflow.
}
\end{figure}

\end{document}